\documentstyle[aps,prl,multicol]{revtex}
\input {epsf.sty}
\begin{document}
\draft
\title{ Charges and Currents in the Noncommutative Chern-Simons Theory of the QHE }
\author{T. H. Hansson and A. Karlhede}
\address{Department of Physics, University of
Stockholm, \\Stockholm Center for Physics, Astronomy and Biotechnology\\ 
S-11385 Stockholm, Sweden }

\date{\today}
\maketitle
\begin{abstract}
We couple the noncommutative 
Chern-Simons theory describing the fractional quantum Hall effect to 
external magnetic and electric potentials, and derive expressions for charge and 
current densities. To lowest non-trivial order the density operator satisfies 
the commutator algebra characteristic of  the 
lowest Landau level. We take this as a  strong indication that the theory 
correctly describes point particles in a strong magnetic field. We 
also obtain the correct QH response to a constant  electric 
field, and density modulations in response to a weak variable magnetic 
field. 
\end{abstract}

%
\newcommand{\ee}{\end{eqnarray}}

\newcommand\ie {{\it i.e. }}
\newcommand\eg {{\it e.g. }}
\newcommand\etc{{\it etc. }}
\newcommand\cf {{\it cf.  }}
\newcommand\grad{\nabla}
\newcommand\half{\frac 1 2 }

\newcommand{\dd}[2]{{\rmd{#1}\over\rmd{#2}}}
\newcommand{\pdd}[2]{{\partial{#1}\over\partial{#2}}}
\newcommand{\pa}[1]{\partial_{#1}}
\newcommand{\pref}[1]{(\ref{#1})}

\newcommand\noi{\noindent}
\newcommand\kp{\ell^{2}(\vec k \times \vec p)}
\newcommand\kph{\frac {\ell^{2}} 2 (\vec k \times \vec p)}

\newcommand {\be}[1]{
       \begin{eqnarray} \mbox{$\label{#1}$}  }


It is well established that the  low-energy effective lagrangian for a 
quantum Hall system is a Chern-Simons (CS) gauge theory. Specifically, for the 
simplest case of the Laughlin fractions, it takes the form,
\be{eff}
{\cal L} = \frac k {4\pi} \epsilon^{\mu\nu\lambda} 
a_{\mu}\partial_{\nu}a_{\lambda} - \frac e {2\pi} \epsilon^{\mu\nu\lambda} 
A_{\mu}\partial_{\nu}a_{\lambda} \, ,
\ee
where $k$ is an odd integer, and $A_{\mu}$ is the external 
electromagnetic potential. This lagrangian (and its generalizations to 
other fractions,  multi-component systems {\em etc.}) correctly describes 
the quantum Hall conductance, and  also 
ground state degenerecies on higher genus surfaces and edge excitations\cite{zee}. 
At the level of the effective theory, the quantization 
of $k$ does not follow from any general 
principle but should be regarded as an input. There are, however, microscopic 
derivations of \pref{eff}, relying on various mean field 
approximations, where the quantization of $k$ essentially follows from 
a statistical transmutation\cite{mf}. 

Another limitation of \pref{eff} follows upon the realization that 
the physics relevant for the Laughlin fractions of the QHE can be   
understood entirely in terms of states in the lowest Landau level 
(LLL).  In particular, this means that the Fourier components of the 
 density operator  do not commute, but obey
\be{ccom}
[\rho_{\vec k}, \rho_{\vec p}]  = - 2i \sin\left(\kph 
\right)e^{\frac {\ell^{2}} 2 \vec k 
\cdot \vec p } \rho_{\vec k+\vec p} \, ,
\ee
where $\ell = (\hbar/eB)^{1/2}$ is the magnetic length\cite{givin}. 
In the effective theory \pref{eff}, however,  $\rho(x) = 2\pi \epsilon^{ij}\pa i 
a_{j} = 2\pi b$, and the commutator \pref{ccom} is $\sim (\vec k\times 
\vec p)\delta(\vec k+\vec p)$.

Recently, Susskind proposed  a noncommutative version of \pref{eff},
\be{moyal}
{\cal L}_{nc} = \frac k {4\pi} \epsilon^{\mu\nu\lambda} \left(
\hat a_{\mu}\star\partial_{\nu}\hat a_{\lambda} + \frac {2i} 3 \hat 
a_{\mu}\star\hat a_{\nu}\star\hat a_{\lambda} \right) \, ,
\ee
where the noncommutative parameter  entering the Moyal star product is 
 $\theta = 1/2\pi\rho$ with $\rho$ being a constant 
background density\cite{suss1}. Susskind argues that 
 $\theta$ can be interpreted as the unit area per particle, 
and that \pref{moyal} incorporates the particle substructure into 
the  effective theory. That this is an essential point when constructing 
effective actions was emphasized by Haldane in the case of one 
dimensional quantum fluids\cite{haldane}. For further discussion of 
the noncommutative CS (NCCS) theory we refer to 
\cite{suss1,poly1,poly2,mor,heller,dimi}.  Here we just mention 
that invariance under large gauge transformations requires 
$k$ to be an integer independent of 
the physical interpretation of the NCCS action\cite{nair1,poly1},
while the restriction to odd integers is a consequence of the 
electrons forming the QH fluid being fermions\cite{suss1}.

In this letter we  extend the  NCCS theory to include 
coupling to an external electromagnetic field. We  then derive 
expressions for charge and current densities, verify the commutation 
relation \pref{ccom} to $O(\hbar\theta^{0})$ in an expansion in $\hbar$ and 
$\theta$. We also show that a weak variable magnetic field gives rise
to density fluctuations, and that the response to a constant 
electric field is given by  the Hall 
conductance $\sigma_{ab}= \sigma_{H} \epsilon_{ab}$. 
These results lend further support to the 
notion that the  NCCS theory is the correct low energy 
theory of the QHE. 

There is an equivalent formulation of NCCS theory in terms of a 
matrix model\cite{poly3}.  To include couplings to external 
 potentials, we propose the following extension of  the matrix version     
 of  Susskind's lagrangian,
\be{new}
L = \frac {eB} 2 {\rm Tr} \{ \left( \dot{ X^{a}}  - i[X^{a}, 
\hat a_{0}]_{m} \right) \left(\epsilon_{ab}X^{b} + \frac 2 {eB} \hat 
A_{a}\right) + 2\theta \hat a_{0} - \frac 2 {eB} \hat A_{0} \} \, ,
\ee
where $X^{a}(t)$ are hermitian matrices, and $\hat a_{0}(t)$ is a 
matrix Lagrange multiplier imposing the (classical) 
commutation relation
\be{tvang}
[X^{a},\epsilon_{ab}X^{b} + \frac 2 {eB} \hat A_{a} ]_{m}=    2i\theta \, .
\ee
The new terms are the couplings to the matrices, 
$\hat A_{\mu}(X^{a},t)$ ($\mu =0,1,2$; $a,b = 1,2$), 
which are related to the ordinary gauge potential, $A_{\mu}(x^{a},t)$,  by the 
Weyl ordering formula,
\be{weyl}
\hat f(X^{a},t) = \int \frac {d^{2}k d^{2}x} {(2\pi)^{2}} \, 
f(x^{a},t) 
  e^{ik_{a}(x-X)^{a}} \, ,
\ee
where $f(x^{a},t)$ is an ordinary function of the usual (commuting) 
coordinate $x^{a}$. To exhibit the canonical structure of the theory, 
we choose a linear gauge, $\hat A_{2} = 0$ and rewrite the constraint 
\pref{tvang} 
as,
\be{ntvang} 
[X^{1}, X^{2} + \frac 1 B \hat A_{1} ]_{m}= 
[X^{1}, \frac 1 B P^{1} ]_{m}= i\theta \, ,
\ee
where $P^{1}$ is the momentum conjugate to $X^{1}$ as calculated from 
the lagrangian \pref{new}. 
For $\hat A$ corresponding to a constant external magnetic field, we 
get $[X^{1},X^{2}]={\rm const.}$.

In the $\hat a_{0}=0$ gauge, the 
lagrangian \pref{new} is quantized  by imposing  equal time
(quantum) commutation relations,
\be{qcom}
[X^{1}_{mn}, P^{1}_{rs}] = i\hbar \delta_{ms}\delta_{nr} \, .
\ee
Note that expressed in canonical variables,
the constraint takes the same form as in the $A_{\mu}=0$ case. Thus the 
whole discussion about constraints, exchange phases etc. will be the 
same as in refs. \cite{suss1} and \cite{poly1}. 
  
We now proceed to analyze the proposed lagrangian \pref{new}.
The gauge transformation $\delta A_{\mu}(x) = 
\partial_{\mu}\Lambda(x)$ implies the following transformation for the 
corresponding matrices,
\be{emgt}
\delta \hat A^{a}_{mn} &=& 
 - \left(\frac {\partial \hat \Lambda }
   {\partial X^{a}}\right)_{mn} = 
   - \frac {\partial {\rm Tr} \hat \Lambda} {\partial X^{a}_{nm}} \\
\delta \hat A_{0} &=& \pa 0 \hat \Lambda \, , \nonumber
\ee
where again  the matrix valued function $\hat\Lambda (X^{a},t)$ is related to 
$\Lambda(x^{a},t)$ by the Weyl ordering \pref{weyl}, and the derivative  
 satisfies  $\partial e^{-ik_{b}X^{b}}/\partial X^{a} = 
 -ik_{a}e^{-ik_{b}X^{b}}$.
It is straightforward to 
show that the lagrangian \pref{new} is invariant both under the `background' 
gauge transformation \pref{emgt}, and the noncommutative gauge 
transformation,
\be{ncgt}
\delta X^{a} &=& i[\hat\lambda, X^{a}]_{m}  \\
\delta \hat a_{0} &=& i[\hat\lambda , \hat a_{0} ]_{m} 
+ \pa{0}\hat\lambda \, .   \nonumber
\ee 
Inserting the proper Weyl ordered expression for $\hat A$ into 
\pref{new} and taking the variation of the corresponding action {\em 
w.r.t.} $A_{\mu}$, 
we obtain the following expressions for the Fourier components of the 
particle  and current densities,
\be{cc}
\rho_{\vec k} &=& {\rm Tr}\left(e^{-ik_{a}X^{a}}\right) \\
j_{\vec k}^{a} &=& {\rm Tr}\left( 
\dot{X^{a}}e^{-ik_{b}X^{b}}\right) \, , \nonumber
\ee
which clearly are invariant under both \pref{emgt} and \pref{ncgt}. 
Note that the 
simple commutation relation \pref{ntvang} is between $X^{1}$ and 
$P^{1}=eBX^{2}+\hat A_{1}$,  not between $X^{1}$ and $X^{2}$ -  
this makes it hard to evaluate  the densities \pref{cc}. 

A noncommutative 
gauge theory does not allow for local observables, and the expressions 
\pref{cc} are in fact closely related to the so called open Wilson 
loops\cite{doug}.
Note that $\rho_{\vec k}$ is the natural noncommutative generalization 
of the usual expression
\be{dens}
\tilde\rho_{\vec k} = \int d^{2}r\,e^{-i\vec 
k \cdot\vec r} \rho(\vec r) =  \int d^{2}r\,e^{-i\vec 
k \cdot\vec r}
\sum_{i=1}^{N} \delta^{2}(\vec r -\vec r_{i}) 
 =   \sum_{i=1}^{N} e^{-i\vec k \cdot\vec r_{i}}  \, .
\ee

The connection between 
the commutative and noncommutative case can be made more precise in 
the modified theory  due to Polychronakos, where the 
constraint \pref{tvang} is relaxed in a minimal way in order to allow for a finite 
system of $N$ particles described by $N\times N$ matrices\cite{poly1}. Up to 
finite size corrections, which should vanish in the large $N$ limit, we 
still expect   the current and charge densities to be given by 
\pref{cc}. In the finite $N$ formulation one can pick a gauge where 
the $2N$ physical degrees of freedom are the diagonal elements of $X^{a}$. 
One of the matrices,  $X^{1}$ say, can be diagonalized, while the off 
diagonal elements of the other are given by 
$X^{2}_{mn}=i\theta/(x_{m}-x_{n})$, where $x_{n}$ are the diagonal 
elements of $X^{1}$. In the limit of large $x^{a}$ and 
large separation compared to $\sqrt\theta$, both matrices are almost 
diagonal and the Fourier transform of $\rho_{\vec k}$ in \pref{cc} 
reproduces $\rho(\vec r) $ in \pref{dens} with 
$\vec r_{n} = (x_{n},y_{n})$. This confirms the  interpretation of 
the diagonal elements of $X^{a}$ as the  positions of the electrons.

The densities \pref{cc} are not local, 
and one would hope that this non-locality would reflect the LLL physics 
of the QH effect. We can now test this proposition
by evaluating the commutator \pref{ccom} using our
density operator \pref{cc}.
This calculation is cumbersome, since it involves 
commutators of exponentials of matrices where the elements do not 
commute. However, using series expansions it is relatively straightforward to obtain
\be{nccc}
[\rho_{\vec k}, \rho_{\vec p}]  = -i \kp \rho_{\vec k+\vec p} + 
O(\hbar\theta , \hbar^{2}) \, ,
\ee
which we recognize as  the leading term in  an $\hbar$ expansion of  \pref{ccom}.
(The sign is related to the sign of $B$ via \pref{qcom}, and 
we follow the convention of refs. \cite{suss1} and \cite{poly1}.)

Note that already to this order the (classical) matrix structure must be 
taken into account in order to reproduce the $\rho_{\vec k +\vec p}$ 
term characteristic for the LLL projection, but also note that
the result is insensitive to the ordering of the 
exponential in \pref{cc}, so to test the Weyl ordering prescription
one must go to higher orders in $\theta$. 
The lagrangian \pref{new}, the particle and current density operators \pref{cc},
and the  commutator \pref{nccc} are the main results of 
this paper.

It is in general hard to calculate the response to an external 
magnetic field since \pref{weyl} only gives an implicit relation for 
the matrix $\hat A^{a}(X^{1},P^{1})$. It can however be done for a weak perturbing 
field of the form $\delta B(\vec x) = \epsilon B \sin (\vec q\cdot\vec x)$, 
where $\epsilon \ll 1$. Substituting this in \pref{weyl} and 
\pref{cc}, and expanding to lowest order in $\epsilon$, the trace can 
be evaluated to lowest order in $\theta$ using a 
coherent state basis, to yield 
\be{densresp}
\rho(\vec x) = \frac 1 {2\pi\theta} \left[ 1 + 
\epsilon \sin (\vec q\cdot\vec x) \right] + O(\theta^{0}) \, 
\ee
as expected.

Next we calculate the current response to an applied constant electric 
field of the form, 
$A_{0} = E_{a}x^{a}$. 
To evaluate \pref{cc}, we first need to determine $\dot {X^{a} }$ from 
the equations of motion, a problem  very similar to the QH 
droplet in a quadratic matrix potential
considered by Polychronakos\cite{poly1}.
Obtaining $\hat A_{0}$ from \pref{weyl}, substituting
in \pref{new} and varying with respect to $X^{a}$, yields 
\be{eom}
B \dot {X^{a}} = -\epsilon_{ab} \hat E^{b} \, .
\ee
Substituting this into the expression \pref{cc} gives the following
expression for the electromagnetic current density $J^{a} = ej^{a}$,
\be{qhe}
J^{a}_{\vec k} = - \epsilon_{ab} \frac {eE^{b}} B \rho_{\vec k} \, ,
\ee
and taking the Fourier transform we finally get 
\be{qhe2}
J^{a} = - \nu \frac {e^{2}} h \epsilon_{ab} E^{b} = 
-\sigma_{H} \epsilon_{ab} E^{b} \, ,
\ee
where the filling fraction $\nu$ is defined by $\rho  = 
\nu eB/h$. From ref. \cite{suss1} we know that for the particles 
described by the matrix model to be fermions, we must have $\nu^{-1} 
=2n +1$ for which \pref{qhe2} gives the Laughlin values 
for the quantized Hall conductance $\sigma_{H}$.

To conclude,  we have presented substantial reasons to believe that  
the noncommutative Chern-Simons Lagrangian \pref{new} correctly 
describes the coupling of a Laughlin  QH liquid to  external 
electromagnetic potentials. That the density commutator \pref{nccc}, 
characteristic of the lowest Landau level, emerges from our 
construction provides further evidence that the theory 
in fact describes point particles in a strong magnetic field. 

It is an obvious and interesting problem to 
evaluate the commutator \pref{nccc}, 
and the density response \pref{densresp} to higher order in 
in $\hbar$ and $\theta$. 
Such calculations should certainly be done in order to see if  
the full $\vec k$ and $\vec p$ dependence in \pref{ccom} is 
reproduced by the operators \pref{cc}. This will show whether the 
full LLL physics is hidden in \pref{new}, or only the leading long 
distance part of it. That there
exists a  mapping from the NCCS theory to the Calogero-Sutherland 
model\cite{poly1}, and also a  connection to the Laughlin 
wavefunctions\cite{heller}, suggest the former scenario, although 
the mapping might be quite complicated\cite{dimi}.
Finally it should  be straightforward to modify our construction to the 
finite droplet  considered in ref. \cite{poly1}. 
This would hopefully
allow for a more detailed analysis of edge modes and edge currents. 

\vskip 2mm\noi
{\bf Acknowledgement:} We thank Alexios Polychronakos for pointing out 
errors in the first version of this paper, and for helping us to get 
the correct form of eq. \pref{emgt}.

\end{document}